\documentclass[a4paper,12pt]{article}

\usepackage[english]{babel}
\usepackage[utf8x]{inputenc}
\usepackage[T1]{fontenc}

\usepackage[a4paper,top=3cm,bottom=2cm,left=3cm,right=3cm,marginparwidth=2cm]{geometry}

\usepackage{setspace}
\usepackage{amssymb}

\usepackage{cite}

\usepackage{amsmath}
\usepackage{graphicx}
\usepackage[colorinlistoftodos]{todonotes}
\usepackage[colorlinks=true, allcolors=blue]{hyperref}

\title{Polarization-independent resonant lattice Kerker effect in phase-change metasurface}

\author{Lei Xiong$^{1,2}$, Xiaoqing Luo$^{2}$, Hongwei Ding$^{1,*}$, Yuanfu Lu$^{2}$, Guangyuan Li$^{2,3,*}$}

\date{}
\begin{document}
\maketitle

\begin{spacing}{2.0}

\noindent $^1$School of Information Science and Engineering, Yunnan University, Kunming 650500, China

\noindent  $^2$CAS Key Laboratory of Human-Machine Intelligence-Synergy Systems, Shenzhen Institute of Advanced Technology, Chinese Academy of Sciences, Shenzhen 518055, China

\noindent $^3$SIAT Branch, Shenzhen Institute of Artificial Intelligence and Robotics for Society, Shenzhen 518055 China


\noindent *Corresponding authors: dhw1964@163.com; gy.li@siat.ac.cn

\end{spacing}

\newpage

\begin{abstract}
Resonant lattice Kerker effect in periodic resonators is one of the most interesting generalizations of the Kerker effect that relates to various fascinating functionalities such as scattering management and Huygens metasurfaces. However, so far this effect has been shown to be sensitive to the incident polarization, restricting its applications. Here, we report, for the first time, polarization-independent resonant lattice Kerker effect in metasurfaces composed of periodic Ge$_2$Se$_2$Te$_5$ (GST) disks. For such a metasurface of square lattice, the spectrally overlap of the electric dipole and magnetic dipole surface lattice resonances can be realized by choosing an appropriate GST crystalline fraction regardless of the incident polarization. The operation wavelength and the required GST crystalline fraction can be conveniently tuned over large ranges since these parameters scale linearly with the disk size and the lattice period, greatly facilitating the design. Making use of the obtained resonant lattice Kerker effect, we realize a reconfigurable and polarization-independent lattice Huygens' metasurface with a dynamic phase modulation of close to $2\pi$ and high transmittance. This work will advance the engineering of the resonant lattice Kerker effect and promote its applications in phase modulation and wavefront control.
\end{abstract}

\newpage

\section{Introduction}
High-index all-dielectric nanoresonators with low losses and multipolar resonances have emerged as new building blocks to realize unique functionalities and novel photonic devices, and have been expected to complement or even replace the plasmonic counterparts \cite{Sci2016Yuri_AllDRev,ACSP2017Yuri_MieRev,NP2017Schilling_MieRev}. Besides the coexistence of strong electric and magnetic resonances, accompanied with resonant enhancement of the electric and magnetic fields, respectively, the interference effects of multipolar resonances have been inspiring intense research efforts since these effects lead to many intriguing phenomena controlling the far-field scattering \cite{Sci2016Yuri_AllDRev,ACSP2017Yuri_MieRev,NP2017Schilling_MieRev,PTRSA2017Yuri_MieIntfRev,NanoP2020Takahara_MieSLRrev}. Among the multipolar interference effects, the destructive interference between the electric and magnetic resonances has been of the focus, since it is relatively simple and it enables unidirectional scattering fulfilling generalized Kerker conditions \cite{NC2013Luk_KerkSiNPradius,ACSNano2013Yuri_KerkSiNPradius,OL2015Alaee_GKerker,OE2018Yuri_KerkerRev,PRL2019_GKerker,LPR2017Babicheva_LatticeKerker,NanoP2018Babicheva_LatticeKerkerSiRod,OL2018Babicheva_LatticeKerkerT,OME2020_LatticeKerkerPerfectRef,OE2021Magnusson_MieKerker}, Huygens' metasurfaces with full phase coverage and high transmission efficiency \cite{AOM2015Yuri_Huygens,LPR2015Boris_Huygens,PRM2020KerkerHuygens}, generalized Brewster effect \cite{NC2016_GBrewster,OE2018_GBrewster,PRAppl2021_GBrewster}, and polarization conversion \cite{APLPhoton2016_KerkPolConv,OE2018_KerkPolConv,OE2019_KerkPolConv}.

In order to achieve the spectral overlap and thus the destructive interference of optically induced electric dipole resonance (EDR) and magnetic dipole resonance (MDR), one needs to tune their spectral positions. This can be done by changing the resonator's parameters, which include the radius \cite{NC2013Luk_KerkSiNPradius,ACSNano2013Yuri_KerkSiNPradius,AOM2015Yuri_Huygens,LPR2015Boris_Huygens}, the height \cite{PRB2017Evlyukhin_KerkHeight}, and the gap of two split dielectric resonators \cite{NL2017Igal_KerkGap}. By further combining free-carrier refraction \cite{ACSP2015Schuller_KerkNtune}, phase-change materials such as germanium antimony telluride (or GST for short) \cite{OL2016Rockstuhl_KerkSNGST,ACSP2019_Non2SupScatPCM,AFM2020Altug_KerkSNGST}, or epsilon-near-zero materials such as indium tin oxide (ITO) \cite{Optica2018Valentine_KerkSiITOtune}, dynamically tunable overlap of the EDR and the MDR has been demonstrated, resulting in programmable Huygens’ metasurfaces and reconfigurable radiation patterns. 

Recently, Mie surface lattice resonances (Mie-SLRs), owing to the coherent interference between the localized Mie resonances and the Rayleigh anomaly (RA) diffracted light, have attracted increasing attention because of their appealing merits including narrow linewidths, large quality factors, strong field enhancements extended over large volumes, and large wavelength tunability \cite{JAP2019Rivas_MieSLR,NanoP2020Takahara_MieSLRrev,RevPhys2021Rasskazov_SLRrev,JAP2021Babicheva_MieSLRrev}. More interestingly, it has been found that the electric dipole and magnetic dipole SLRs (ED-SLR and MD-SLR) can be tuned independently to certain wavelengths by choosing the array periodicity in a different direction \cite{LPR2017Babicheva_LatticeKerker}, facilitating the realization of the resonance overlap. The Kerker effect due to the destructive interference between the ED-SLR (or the MD-SLR) of the array and the MDR (or the EDR) of single nanoparticles, or between the ED-SLR and the MD-SLR, was referred to as the {\sl resonant lattice Kerker effect} by Babicheva and Evlyukhin \cite{LPR2017Babicheva_LatticeKerker}. This effect has been regarded as the most interesting generalization of the Kerker effect since only the zeroth-order diffraction is involved \cite{OE2018Yuri_KerkerRev}. In other words, the resulting forward-propagating beam of the resonant lattice Kerker effect does not spread in side directions. This is favorable in many applications compared with the conventional Kerker effect due to the EDR and MDR overlap of single nanoparticles, for which a larger portion of the light is scattered in the side directions \cite{LPR2017Babicheva_LatticeKerker}. 

By varying one of the lattice period, resonant lattice Kerker effect due to the destructive interference of the ED-SLR of the array and the MDR of single nanoparticles can be implemented, resulting in resonant suppression of light reflection \cite{LPR2017Babicheva_LatticeKerker,NanoP2018Babicheva_LatticeKerkerSiRod} or transmission \cite{OL2018Babicheva_LatticeKerkerT}, perfect reflection \cite{OME2020_LatticeKerkerPerfectRef} or perfect absorption \cite{ACSP2018_LatticeKerkerPerfectAbs,OE2020_LatticeKerkerPerfectAbs,OL2021_LatticeKerkerPerfectAbs}. Quite recently, some of the authors \cite{JPD2022Li_KerkSLRGST} made use of GST metasurfaces and achieved active tuning of the ED-SLR and MDR overlap by changing the GST crystalline fraction, which quantifies the fraction of crystallized material during GST experiences nonvolatile transition from the amorphous state to the crystalline state. This leads to the transition between the ED-SLR and the resonant lattice Kerker effect, enabling multilevel tuning of reflection, transmission and absorption with large modulation depths. However, as the linear polarization of incident light changes, the transition between the ED-SLR and the MD-SLR, or between the MDR and the EDR can be observed, resulting in that the MD-SLR and the EDR cannot overlap with each other, as shown by ref~\cite{LPR2017Babicheva_LatticeKerker}. As a result, the resonant lattice Kerker effect due to the ED-SLR and MDR overlap, or the MD-SLR and EDR overlap, is sensitive to the incident polarization, restricting its potential applications.

Taking advantage of the transition between the ED-SLR and the MD-SLR upon the change of the incident polarization, it is possible to achieve resonant lattice Kerker effect under the same conditions for two orthogonal linear polarizations based on the destructive interference of the ED-SLR and the MD-SLR in a square lattice. However, by illustrating with two types of arrays that consist of either silicon nanoparticles or core-shell nanoparticles, Babicheva and Evlyukhin \cite{LPR2017Babicheva_LatticeKerker} found that the ED-SLR and the MD-SLR are spectrally close but do not overlap when lattice periods in the $x$ and $y$ directions are equal, that is, $\Lambda_x=\Lambda_y$, and that $\Lambda_x$ and $\Lambda_y$ should be different in order to achieve the resonant lattice Kerker effect. As a consequence, the obtained resonant lattice Kerker effect due to the ED-SLR and MD-SLR overlap remains sensitive to the incident polarization because of the unequal lattice periods. Additionally, even when $\Lambda_x=\Lambda_y$, the resonances for two linear polarizations that are not orthogonal (for instance, with polarization angles are 0$^\circ$ and 30$^\circ$) may be different, since the nanodisks pattern in these directions are different. As a result, the conditions for achieving the resonant lattice Kerker effect would also be different. In other words, to date all the reported resonant lattice Kerker effects have been polarization dependent, which is distinct from the conventional Kerker effect without involving the SLR(s), and it remains challenging to realize polarization-independent resonant lattice Kerker effect.


\begin{figure}[!hbt]
\centering
\includegraphics[width=83mm]{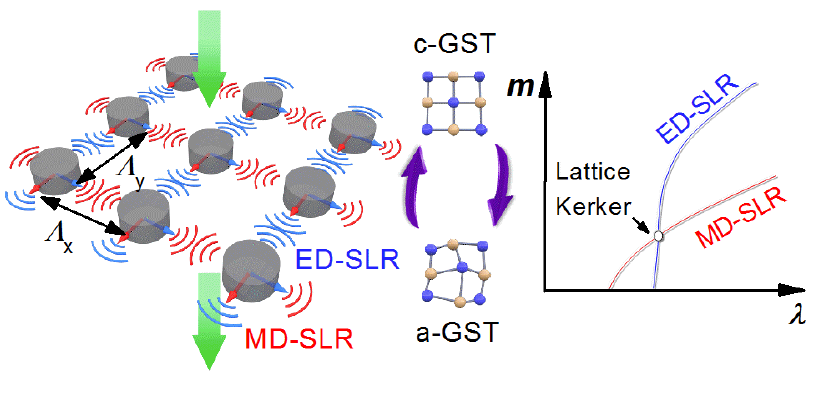}
\caption{Schematics of the GST metasurface composed of periodic disks with $\Lambda_x=\Lambda_y=\Lambda$ supporting polarization-independent resonant lattice Kerker effect. The ED-SLR and MD-SLR overlap is enabled by changing the GST crystalline.}
\label{fig:schem}
\end{figure}

In this work, we tackle this challenge and numerically demonstrate, for the first time, polarization-independent resonant lattice Kerker effect based on the destructive interference of the ED-SLR and the MD-SLR in metasurfaces composed of Ge$_2$Se$_2$Te$_5$ (GST) disk array with equal lattice periods ($\Lambda_x=\Lambda_y=\Lambda$). As illustrated by figure~\ref{fig:schem}, the ED-SLR and MD-SLR overlap is enabled by choosing the GST crystalline fraction. We will show that for any incident polarization, including linear polarizations with different polarization angles, circular and elliptical polarizations, the far-field reflectance/transmittance spectra and the ED-SLR and MD-SLR overlap remain unchanged, although the near-field distributions of the ED-SLR and the MD-SLR are sensitive to the polarization. The requirements for achieving this polarization-independent characteristic will be clarified. We will also show that the operation wavelength and the GST crystalline fraction required for the ED-SLR and MD-SLR overlap can be linearly tuned by varying the geometric size. We will further illustrate the application by numerically demonstrating polarization-independent reconfigurable Huygens’ metasurface with full transmission phase coverage of 2$\pi$ and dynamic phase modulation up to $0.8\cdot 2\pi$ with high transmittance. This work points to new approaches for realizing resonant lattice Kerker effect and reconfigurable Huygens' metasurfaces. 

\section{Design}
Figure~\ref{fig:schem} illustrates the designed metasurface composed of GST disks arranged in the square array with lattice period $\Lambda$ in both $x$ and $y$ directions. These disks have diameter $d$ and height $h$. The metasurface is illuminated by linearly polarized plane wave light under normal incidence. 

The far-field zeroth-order reflection and transmission coefficient spectra, $r_0(\lambda)$ and $t_0(\lambda)$, as well the near-field distributions of the designed metasurface were numerically calculated with a home-developed package for the fully vectorial rigorous coupled-wave analysis (RCWA), which was developed following refs \cite{JOSAA1995RCWA,JOSAA1997RCWA,PRB2006RCWA}. The zeroth-order reflectance and transmittance spectra are then obtained with $R_0(\lambda)=|r_0(\lambda)|^2$ and $T_0(\lambda)=|t_0(\lambda)|^2$, and the transmission phase spectra are obtained with $\phi_t(\lambda) = {\rm arg}\{t_0(\lambda)\}$. Wavelength dependent permittivities of the GST material in different crystallization levels are determined by
\begin{eqnarray}
\label{Eq:epsiEff}
\frac{\varepsilon_{\rm eff}(\lambda)-1}{\varepsilon_{\rm eff}(\lambda)+2}=m\frac{\varepsilon _{\rm c}(\lambda)-1}{\varepsilon_{\rm c}(\lambda)+2}+\left ( 1-m \right )\frac{\varepsilon_{\rm a}(\lambda)-1}{\varepsilon_{\rm a}(\lambda)+2} \,,
\end{eqnarray}
where $m$ is the GST crystalline fraction with $0 \leq m \leq 1$, and $\varepsilon_{\rm a}(\lambda)$ and $\varepsilon _{\rm c}(\lambda)$ are the complex permittivities of the amorphous ($m=0$) and crystalline ($m=1$) GST, respectively. The values of $\varepsilon_{\rm a}(\lambda)$ and $\varepsilon _{\rm c}(\lambda)$ are taken from ref~\cite{ProcSPIE2017GSTnk} and those of $\varepsilon_{\rm eff}(\lambda)$ are shown in figure S1, the supplementary document. Unless otherwise specified, the numerical simulations were performed with $h=740$~nm, $d=1~\mu$m, $\Lambda=3~\mu$m, and the incident electric field was set to be polarized along the $x$ direction.

\section{Results and discussion}
\subsection{ED-SLR and MD-SLR overlap enabled by GST's phase change}

\begin{figure*}[!hbt]
\centering
\includegraphics[width=150mm]{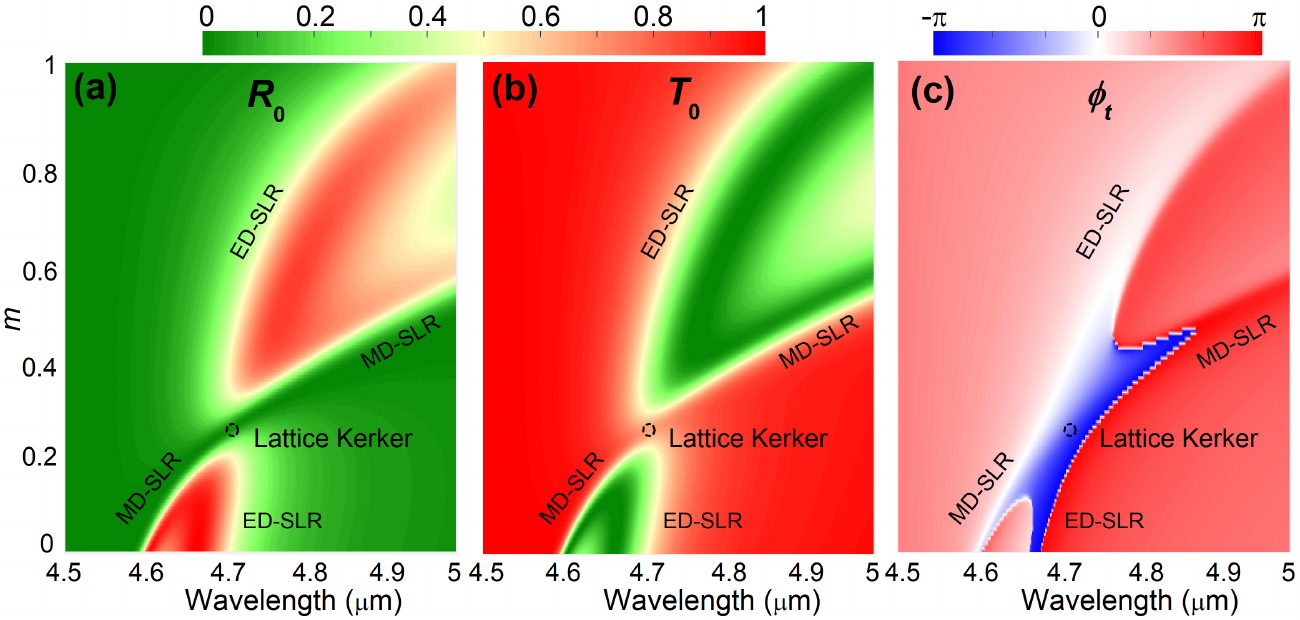}
\caption{Simulated zeroth-order (a) reflectance $R_0$, (b) transmittance $T_0$, and (c) transmission phase $\phi_t$ spectra of the designed GST metasurface with different crystalline fractions. Overlap of ED-SLR and MD-SLR takes place around $m_{\rm Kerker}=0.25$ and $\lambda_{\rm Kerker}=4.71~\mu$m, suggesting the occurrence of the resonant lattice Kerker effect, as indicated by black circles. }
\label{fig:RTvsWvM}
\end{figure*}

Figure~\ref{fig:RTvsWvM}(a)(b) depicts the simulated zeroth-order reflectance and transmittance spectra of the designed GST metasurface with different crystalline fractions $m$. Results show that there exist two branches of resonances, as evidenced by the reflectance peaks and the transmittance dips. For the amorphous GST metasurface ($m=0$), the two branches of resonances are spectrally close but do not overlap, consistent with the results of periodic silicon or core-shell nanoparticles \cite{LPR2017Babicheva_LatticeKerker}. As $m$ increases, these two branches are both red-shifted but to different extents. This results in a crossing or spectral overlap of these two resonances at the wavelength of $\lambda_{\rm Kerker}=4.71~\mu$m for $m_{\rm Kerker}=0.25$, as indicated by the dashed circle. In this scenario, the reflectance approaches zero and the transmittance is high, suggesting the occurrence of the Kerker effect.

\begin{figure*}[!hbt]
\centering
\includegraphics[width=150mm]{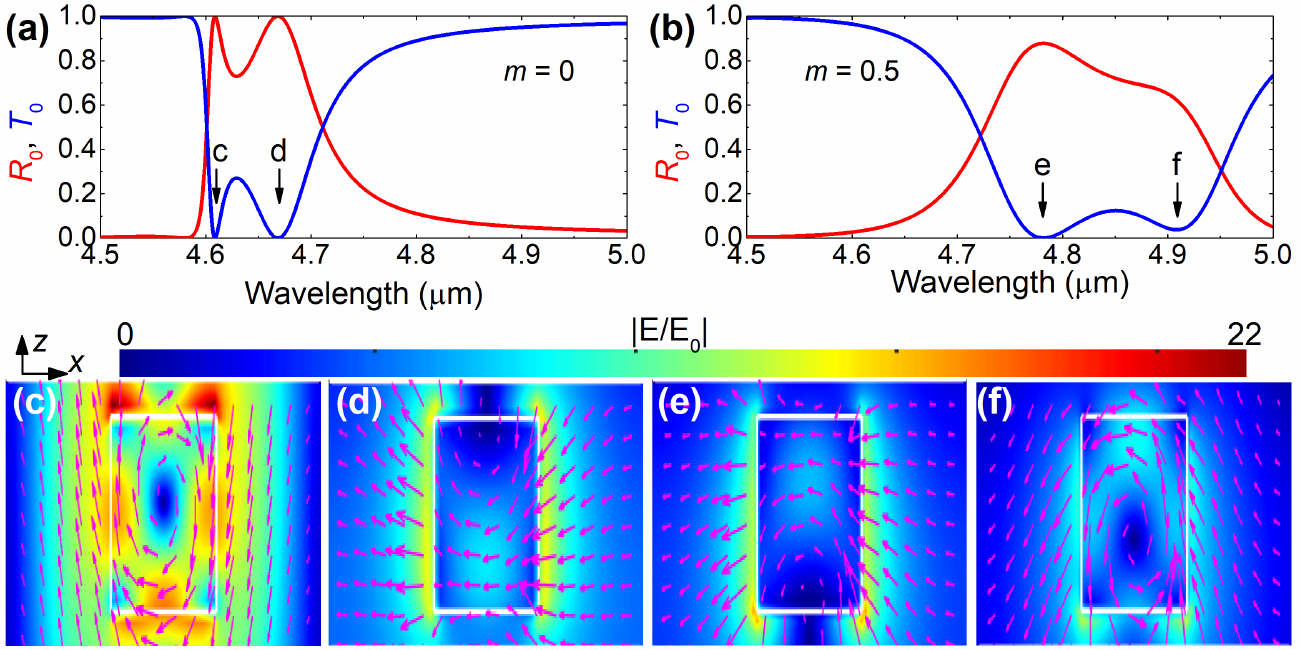}
\caption{(a)(b) Simulated reflectance and transmittance spectra for (a) $m=0$ and (b) $m=0.5$. (c)--(f) Simulated near-field electric field distributions (normalized to the incidence, color for the amplitude and arrows for the directions) in side views at (c) $\lambda=4.61~\mu$m and (d) $\lambda=4.67~\mu$m for $m=0$, and at (e) $\lambda=4.78~\mu$m and (f) $\lambda=4.91~\mu$m for $m=0.5$. }
\label{fig:Fields}
\end{figure*}

In order to understand the underlying physics, in figures~\ref{fig:Fields}(a)(b) we replot the zeroth-order reflectance and transmittance spectra for $m=0$ and 0.5, respectively, and in figure~\ref{fig:Fields}(c)--(f) we plot the near-field electric field distributions of the resonances as suggested by the transmittance dips (or the reflectance peaks). Figure~\ref{fig:Fields}(a) shows that for $m=0$ the two resonances are excited at the wavelengths of $\lambda=4.61~\mu$m and $\lambda=4.67~\mu$m, respectively. At $\lambda=4.61~\mu$m, the electric field distributions show a circulation inside the GST disk, corresponding to a magnetic dipole, and the strongly enhanced electric field extends over large volumes, as shown by figure~\ref{fig:Fields}(c). These are typical characteristics of the MD-SLR \cite{JAP2019Rivas_MieSLR,ACSP2020Rivas_MieSLR,AOM2020Rivas_MieSLR,NanoMat2022Li_SLRGST}. At $\lambda=4.67~\mu$m, figure~\ref{fig:Fields}(d) shows that the electric field is aligned with the incident field and is greatly enhanced and extended over large volumes, confirming the excitation of the ED-SLR \cite{JAP2019Rivas_MieSLR,ACSP2020Rivas_MieSLR,AOM2020Rivas_MieSLR,NanoMat2022Li_SLRGST}.

\begin{figure*}[hbt]
\centering
\includegraphics[width=170mm]{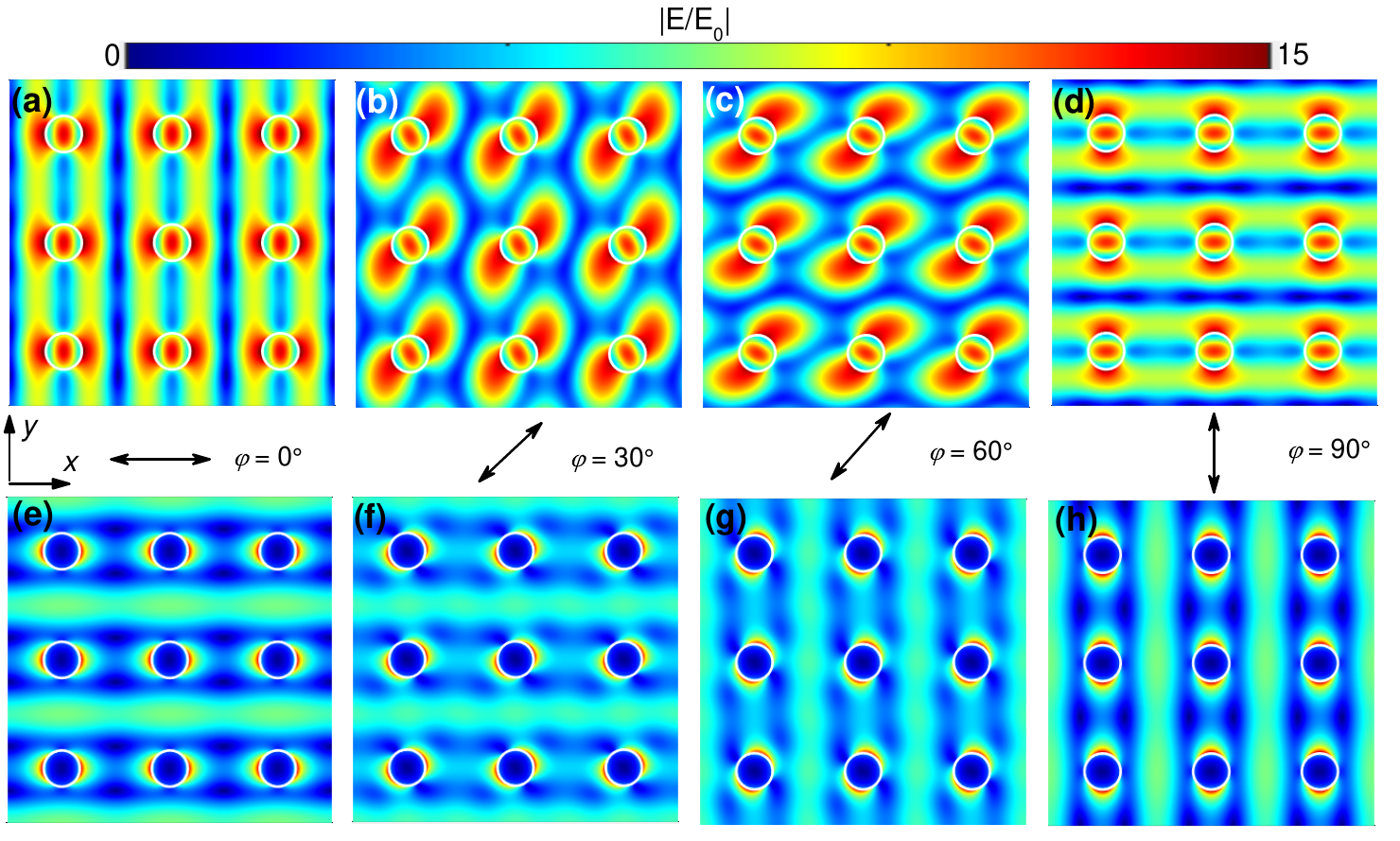}
\caption{Simulated near-field electric field distributions in top views at (a)--(d) $\lambda=4.61~\mu$m and (e)--(h) $\lambda=4.67~\mu$m for $m=0$. From left to right, the incident polarization angles are (a)(e) $\varphi=0^\circ$ ($x$ polarization), (b)(f) $30^\circ$, (c)(g) $60^\circ$, and (d)(h) $90^\circ$ ($y$ polarization).}
\label{fig:FieldsTop}
\end{figure*}

For the metasurface with semi-crystalline GST ($m=0.5$), figure~\ref{fig:Fields}(b) shows that the ED-SLR and the MD-SLR are respectively excited at $\lambda=4.78~\mu$m and $\lambda=4.91~\mu$m. We assign these two resonances as the ED-SLR and the MD-SLR, respectively, because their near-field electric fields, as shown in figures~\ref{fig:Fields}(e)(f), have the same distributions as figures~\ref{fig:Fields}(d)(c), respectively. 

Note that for the two resonances for $m=0$ in figure~\ref{fig:Fields}(a), $R_0+T_0=1$ since the amorphous GST is lossless in this spectral range. However, for the two resonances for $m=0.5$ in figure~\ref{fig:Fields}(b), $R_0+T_0<1$ because the semi-crystalline GST suffers from relatively low absorption loss. Nevertheless, the reflectance reaches zero when the ED-SLR and the MD-SLR are overlapped for $m=0.25$. In other words, the relatively low GST absorption loss does not prevent compensation of electric and magnetic polarizability for achieving the resonant lattice Kerker effect.

Therefore, we have numerically demonstrated that both the ED-SLR and MD-SLR are excited, and that these SLRs can be spectrally tuned by changing the GST crystalline fraction, and that it is possible to realize the ED-SLR and MD-SLR overlap, {\sl i.e.}, the resonant lattice Kerker effect by choosing appropriate GST crystalline fraction.

\subsection{Polarization independence}
Although the above results are obtained for the specific incident polarization along the $x$ direction, we find that the far-field spectra in figure~\ref{fig:RTvsWvM} are independent from the incident polarization, including any linear polarization angle, circular or elliptical polarization (figure S2 and S3, the supplementary document). To understand this striking characteristic, we turn to the Jones calculus. The electric field vector of any polarization state can be expressed based on its orthogonal component $E_x$ and $E_y$,
\begin{eqnarray}
\label{Eq:EJones}
{\sl {\bf E}}= [A] \left[ 
\begin{aligned}
&E_x \\
&E_y
\end{aligned}
\right]\,,
\end{eqnarray}
where $[A]$ is a Jones matrix. If we know the reflectance coefficients of $E_x$ and $E_y$,
\begin{eqnarray}
\label{Eq:RefJones}
\left[ 
\begin{aligned}
&E_{x,{\rm ref}} \\
&E_{y,{\rm ref}}
\end{aligned}
\right]= \left[ 
\begin{aligned}
&r_{xx} \quad  r_{xy} \\
&r_{yx} \quad  r_{yy}
\end{aligned}
\right] \left[ 
\begin{aligned}
&E_{x,{\rm in}} \\
&E_{y,{\rm in}}
\end{aligned}
\right]\,,
\end{eqnarray}
the reflectance coefficient of this polarization state can be expressed as
\begin{eqnarray}
\label{Eq:RefE}
{\sl {\bf E}}_{\rm ref}= r {\sl {\bf E}}_{\rm in}\,,
\end{eqnarray}
provided $r_{xx}= r_{yy}=r$ and $r_{xy}= r_{yx}=0$ (the detailed derivation can be found in the supplementary document). In other words, if the reflectance coefficients for $E_x$ and $E_y$ are equal and there is no polarization conversion between these two orthogonal components, the reflectance coefficient will be polarization independent. This precondition can be satisfied for a metasurface composed of periodic disks and square lattices, regardless of whether SLRs or coupling between disks are involved. Similar derivation and the same conclusion also hold for the transmittance coefficient.

Furthermore, the assignments of these two branches of resonances to the MD-SLR and the ED-SLR remain unchanged. This is illustrated by the near-field electric fields in top views for the two resonances at $\lambda=4.61~\mu$m and $\lambda=4.67~\mu$m for $m=0$, as shown by figure~\ref{fig:FieldsTop}. For the incident polarization that is parallel to the $x$ axis, or $\varphi=0^\circ$, figure~\ref{fig:FieldsTop}(a) shows that a magnetic dipole along the polarization direction and a standing grazing wave (Rayleigh anomaly) in the same direction are excited; whereas figure~\ref{fig:FieldsTop}(e) shows the excitation of a electric dipole along the polarization direction and a standing grazing wave in the perpendicular direction. These corresponding to the MD-SLR and the ED-SLR, respectively. As the incident polarization angle increases to $\varphi=90^\circ$, both the magnetic dipole and the electric dipole rotate accordingly, keeping their directions along the incident polarization. On the other hand, the direction of the standing grazing wave for the MD-SLR switches from the $x$ direction to the $y$ direction, as shown by figures~\ref{fig:FieldsTop}(a)--(d), whereas that for the ED-SLR switches from the $y$ direction to the $x$ direction, as shown by figures~\ref{fig:FieldsTop}(e)--(h).

Therefore, by having periodic disks with  square lattice with $\Lambda_x=\Lambda_y$, we have shown that the resonant lattice Kerker effect enabled by choosing the GST crystalline fraction is independent from the incident polarization in terms of the far-field spectra and the assignments of the resonances, although the near-field distributions are sensitive to the polarization.

\begin{figure*}[!hbt]
\centering
\includegraphics[width=150mm]{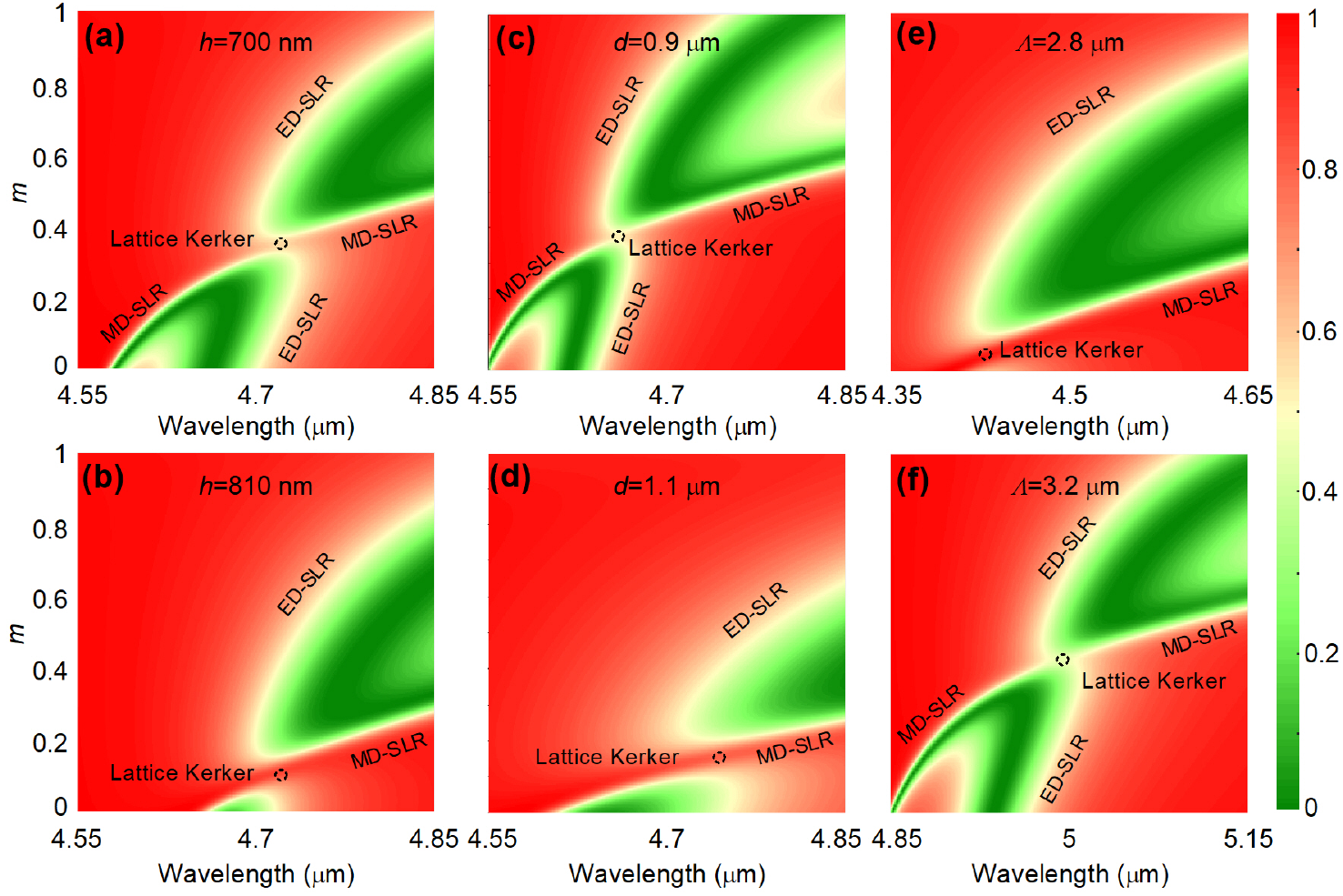}
\caption{Simulated zeroth-order transmittance $T_0$ for (a) $h=700$~nm and (b) $h=810$~nm, (c) $d=0.9~\mu$m and (d) $d=1.1~\mu$m, (e) $\Lambda=2.8~\mu$m and (f) $\Lambda=3.2~\mu$m. Black circles indicate the wavelength and the required GST crystalline fraction for the ED-SLR and MD-SLR overlap, or resonant lattice Kerker effect.}
\label{fig:TvsHDP}
\end{figure*}

\begin{figure*}[!hbt]
\centering
\includegraphics[width=150mm]{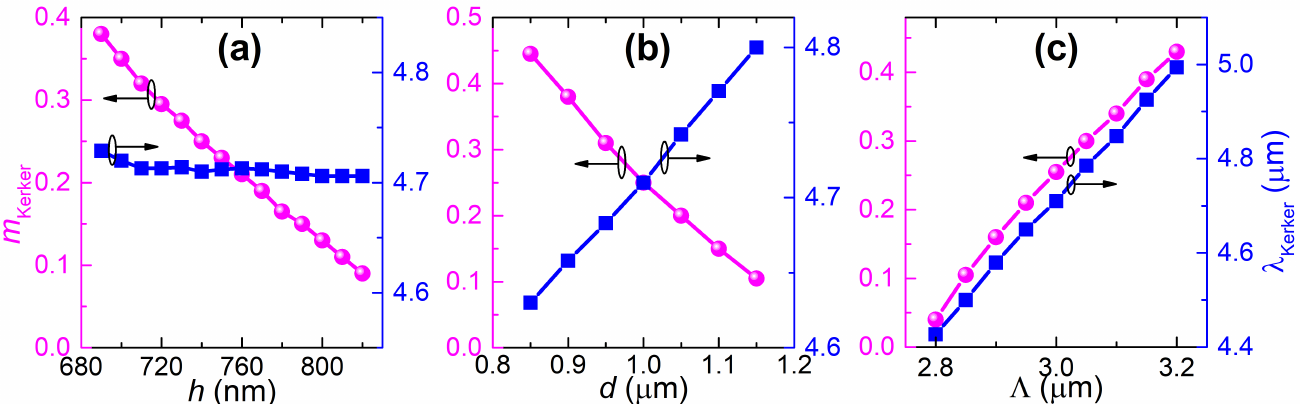}
\caption{Required GST crystalline fractions $m_{\rm Kerker}$ and the operation wavelengths $\lambda_{\rm Kerker}$ of the ED-SLR and MD-SLR overlap as functions of (a) GST disk height, (b) diameter, and (c) lattice period.}
\label{fig:MWvvsHDP}
\end{figure*}

\subsection{Tunability by changing geometric parameters}

Because the ED-SLR (or MD-SLR) is due to the coherent interference between the Mie EDR (or MDR) and the RA diffracted light, the Mie resonances strongly depend on the GST disk size, and the RA wavelength is determined by $n_0 \Lambda$ under normal incidence, we expect that the wavelength as well as the GST crystalline fraction required for achieving the ED-SLR and MD-SLR overlap can be tuned by varying the geometric parameters.

Figures~\ref{fig:TvsHDP}(a)(b) show that, for $h=700$~nm and $h=810$~nm, the lattice Kerker effect takes place at almost the same wavelength as that for $h=740$~nm in figure~\ref{fig:RTvsWvM}. Taller disks require smaller GST crystalline fraction for achieving the resonance overlap. Figure~\ref{fig:MWvvsHDP}(a) depicts the required GST crystalline fractions and the operation wavelengths of the ED-SLR and MD-SLR overlap, $m_{\rm Kerker}$ and $\lambda_{\rm Kerker}$, for different disk heights. It is clear that $m_{\rm Kerker}$ decreases linearly with $h$, whereas $\lambda_{\rm Kerker}$ decreases slightly with $h$. These results suggest that the metasurface composed of taller GST disks requires smaller crystalline change in order to achieve the ED-SLR and MD-SLR overlap that occurs at the same wavelength. 

\begin{figure*}[!hbt]
\centering
\includegraphics[width=150mm]{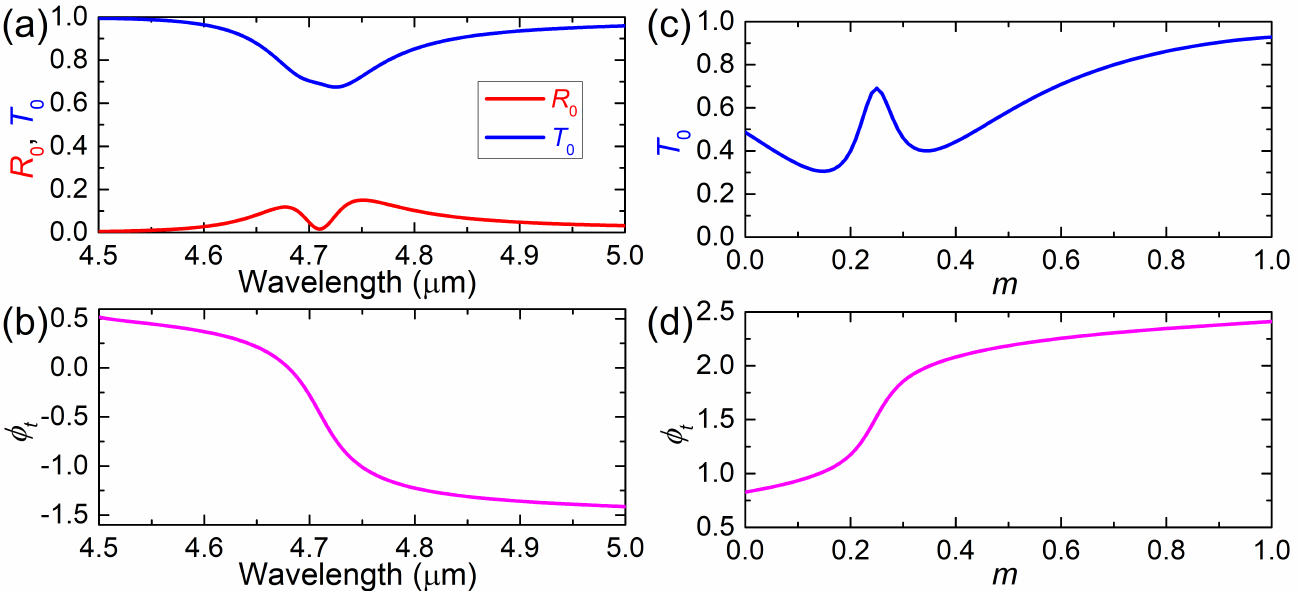}
\caption{(a) Simulated reflectance and transmittance spectra, and (b) the corresponding transmission phase spectra for $m_{\rm Kerker}=0.25$. (c) Simulated transmittance and (d) the corresponding phase at the operating wavelength of $\lambda_{\rm Kerker}=4.71~\mu$m versus GST crystalline fraction.}
\label{fig:Phase}
\end{figure*}

Figures~\ref{fig:TvsHDP}(c)(d) and \ref{fig:RTvsWvM}(b) show that, the larger the GST disk diameter $d$, the smaller the required GST crystalline fraction and the longer the operation wavelength for the ED-SLR and MD-SLR overlap. Indeed, figure~\ref{fig:MWvvsHDP}(b) shows that, as $d$ increases, $m_{\rm Kerker}$ decreases linearly, whereas $\lambda_{\rm Kerker}$ increases linearly. Comparing figures~\ref{fig:TvsHDP}(e)(f) and \ref{fig:RTvsWvM}(b), we find that the larger the lattice period $\Lambda$, the larger the required GST crystalline fraction and the longer the operation wavelength for the occurrence of the ED-SLR and MD-SLR overlap. Figure~\ref{fig:MWvvsHDP}(c) further shows that both $m_{\rm Kerker}$ and $\lambda_{\rm Kerker}$ increases linearly with $\Lambda$.

These behaviors can be understood by the spectral shifts of the ED-SLR and MD-SLR. Since the EDR is independent from $h$, whereas the MDR experiences redshift that linearly scales with $h$, the ED-SLR are almost independent from $h$, whereas the MD-SLR red-shifts linearly with $h$. The smaller spectral distance between the ED-SLR and the MD-SLR requires a smaller GST crystalline fraction in order to achieve the resonance overlap. On the other hand, the wavelength for this overlap decreases slightly. Similarly, as the disk diameter $d$ increases, both the EDR and the MDR are redshifted but to different extends, resulting in redshifted ED-SLR and MD-SLR but with smaller spectral distance. As a consequence, $m_{\rm Kerker}$ decreases whereas $\lambda_{\rm Kerker}$ increases. When the lattice period increases, the ED-SLR and the MD-SLR are both redshifted but with larger spectral distance. Thus, both $m_{\rm Kerker}$ and $\lambda_{\rm Kerker}$ increase. 

Therefore, in order to minimize the GST crystalline fraction $m_{\rm Kerker}$ for achieving the ED-SLR and MD-SLR overlap, for example, to realize $m_{\rm Kerker}=0$ to the extreme limit, one can increase the disk height or diameter, or decrease the lattice period. This points to alternative approach to achieve polarization-independent resonant lattice Kerker effect, especially in periodic silicon or core-shell nanoparticles, of which the refractive index cannot be easily tuned.

\subsection{Reconfigurable and polarization-independent lattice Huygens’ metasurface}
Based on the obtained polarization-independent resonant lattice Kerker effect enabled by changing the GST crystalline fraction, we further realize reconfigurable and polarization-independent {\sl lattice Huygens’ metasurfaces}. Here, ``lattice'' is added in order to highlight the key role of the SLRs, which involve only the zeroth-order diffraction and suppress scattering in side directions, inherited from the resonant lattice Kerker effect.

Figure~\ref{fig:RTvsWvM}(c) depicts the simulated transmission phase spectra of the designed metasurface for different GST crystalline fractions. Results show that a complete phase coverage from 0 to $2\pi$ across the spectral position of the ED-SLR and MD-SLR overlap, accompanied by high transmittance above 68\%, can be achieved when $m_{\rm Kerker}=0.25$, as better visualized by figures.~\ref{fig:Phase}(a)(b). Note that for any incident polarization, the results remain unchanged. These suggest that the designed metasurface with the resonant lattice Kerker effect can function as a polarization-independent lattice Huygens' metasurface. 

By further making use of the intermediate GST crystalline fraction, figure~\ref{fig:RTvsWvM}(b)(c) shows that the obtained polarization-independent lattice Huygens' metasurface can be tuned actively. In figure~\ref{fig:Phase}(c)(d) we replot the simulated transmittance and the associated phase values for different GST crystalline fractions at the wavelength of $\lambda_{\rm Kerker}=4.71~\mu$m, when the ED-SLR and MD-SLR overlap occurs. We find that a dynamic phase modulation of $\Delta \phi=0.8\cdot 2\pi$ can be achieved accompanied with high transmittance of above 31\%. This performance is comparable to that of a Huygens’ metasurface composed of multi-layer Ge/GST nanodisks, which is based on conventional Kerker effect without involving SLR(s) \cite{AFM2020Altug_KerkSNGST}. 

\section{Concluding remarks}
In conclusions, we have theoretically demonstrated polarization-independent resonant lattice Kerker effect in a phase-change metasurface. Simulation results have shown that by properly choosing the GST crystalline fraction, the ED-SLR and the MD-SLR, which initially do not spectrally overlap for the amorphous GST when the lattice periods in the $x$ and $y$ directions are not equal, can be tuned to the spectral overlap. This leads to the polarization-independent resonant lattice Kerker effect. We have attributed the polarization-independent characteristic to the periodic GST disks with square lattice, which have equal reflectance/transmittance coefficients for the $x$ and $y$ polarizations and have no polarization conversion. We have also shown that the operation wavelength can be adjusted linearly by varying the disk diameter and the lattice period, but it is almost independent from the disk height, facilitating the design. On the other hand, the GST crystalline fraction required for achieving the ED-SLR and MD-SLR overlap decreases linearly with the disk height and diameter, but increases linearly with the lattice period. These behaviours point to the alternative approach to realize the polarization-independent resonant lattice Kerker effect: by properly choosing the geometric size for a given the material system. This approach, which is equivalent to the one adopted in this work, that is, changing the GST crystalline fraction for a given geometry, deserves further investigation especially in metasurfaces composed of periodic silicon nanoparticles. 

Taking advantage of the ED-SLR and MD-SLR overlap enabled by changing the GST crystalline fraction, we managed to realize reconfigurable and polarization-independent lattice Huygens' metasurfaces. We have shown that full transmission phase coverage of $2\pi$ and high transmittance above 68\% over the spectral range, and dynamic phase modulation of up to $0.8\cdot 2\pi$ with transmittance above 31\% can be achieved. Therefore, this work also suggests a new approach to realize programmable polarization-independent Huygens' metasurface, which, we expect, would be more favorable over that based on conventional Kerker effect, since the scattering in the side directions is well suppressed thanks to the lattice effect. As a final remark, we expect that the proposed polarization-independent resonant lattice Kerker effect will be demonstrated and find applications in other operation wavelengths by tuning the geometric parameters, and that the transmittance can be significantly improved to near unity by adopting other low-loss or even lossless phase-change materials such as Ge$_2$Sb$_2$Se$_4$Te (GSST) \cite{NC2019_GSST}, Sb$_2$S$_3$ and Sb$_2$Se$_3$ \cite{AFM2020Muskens_SbSSe,NL2021Singh_SbS}.


\section*{Acknowledgments}
This work was supported by the Shenzhen Research Foundation (JCYJ20180507182444250) and the State Key Laboratory of Advanced Optical Communication Systems and Networks, China (2020GZKF004).

\bibliographystyle{unsrt}
\bibliography{sample}

\end{document}